\documentclass[12pt]{elsart}
\usepackage{epsfig}
\usepackage{palatino}

\begin{document}

\begin{frontmatter}
\title{Nuclear Transparency in Heavy Ion Collisions at 14.6 GeV/nucleon}

\author{H.M. Ding \thanksref{Ding}},
\author{P. Gl\"assel}, 
\author{J. H\"ufner\thanksref{Huefner}}
\thanks[Ding]{On leave of absence from Department of Physics, Jilin
University, Changchun, China.}
\thanks[Huefner]{Corresponding Address: Institut f\"ur Theoretische Physik, Universit\"at Heidelberg,
Philosophenweg 19, D-69120 Heidelberg, Germany. Tel: 06221-549440; Fax: 06221-549331;
Email: Joerg.Huefner@urz.uni-heidelberg.de.}

\emph{Fakult\"at f\"ur Physik und Astronomie,} \\
\emph{Universit\"at Heidelberg, Germany}\\[1cm]

{\it submitted to Nuclear Physics A}

\begin{abstract} 

The probability of a projectile nucleon to traverse a target nucleus
without interaction is calculated for central Si-Pb
collisions and compared to the data of E814. The calculations are
performed in two independent ways, via Glauber theory and using the
transport code UrQMD. For central collisions Glauber predictions are
about 30 to 50\% higher than experiment, while the output of UrQMD does
not show the experimental peak of beam rapidity particles.

\textbf{PACS number: 25.75.-q}

\textbf{Keywords: Nuclear Transparency; Wounded Nucleon}
\end{abstract}
\end{frontmatter}

The dynamics of a heavy ion collision is a complicated many-body problem. It
is the task of appropriately designed experiments to isolate one particular
aspect of the dynamics and elucidate its physics. Wounded nucleons are one of
the open problems. In a heavy ion collision a nucleon may undergo a sequence
of collisions, which follow so rapidly one after another, that the nucleon is
no more in its ground state, even not necessarily in any definite excited baryonic
resonance (like $\Delta $ or $N^{*}$). We will speak of a wounded
nucleon. In a next encounter with another nucleon this wounded nucleon will
not interact with the free space NN cross section $\sigma_{in}^{NN}$ but
with an effective one $\sigma_{\rm eff}$. Is it possible to determine $\sigma_{\rm eff}$
from experiment?

The E814 collaboration has designed an experiment which may answer
this question \cite{1}.  At the energy of 14.6 GeV/nucleon, the
projectile $^{28}$Si collides with Al, Cu and Pb and the beam
rapidity nucleons are studied as a function of the centrality of the
reaction (controlled by a measurement of the transverse energy). The
beam-rapidity nucleons belong to the projectile and have not lost any
energy in the reaction. Of course, in a peripheral reaction one always
has the so called spectator nucleons, which never hit the target
nucleus. They are not interesting for our purpose. However, in a
central event, e.g., for Si on Pb, one still sees beam rapidity
nucleons, i.e., nucleons of Si which go through the target nucleus
without energy loss.  The target nucleus is transparent for these
projectile nucleons. We expect the transparency of a heavy nucleus
like Pb to be small. Indeed, the ``survival probability'' $S$ for a
projectile nucleon to pass through the target without any inelastic
interaction has been measured to $S_{exp}=3.5\times 10^{-3}$ for
central Si-Pb collisions \cite{1}. Does this result contain
information about wounded nucleons and effective NN cross sections?

We proceed in the following ways: The transparency is calculated \emph{without}
assuming any exotic phenomena and then the quality of agreement with experiment
is judged. We use two complementary methods of calculation

\begin{itemize}

\item Glauber theory, in which the physics and formalism is simple and
transparent, though not always realistic;

\item Cascade calculation (UrQMD), which represents present-day
state-of-the-art and may be compared to earlier calculations (Bass 
et al. \cite{2}).

\end{itemize}

The survival probability $S$ for a projectile nucleon to
pass through the target without inelastic interaction is calculated in a
Glauber-type approximation (straight trajectories, frozen nucleons) as

\begin{equation}
S_{G}(b,\sigma_{\rm eff})= \int \d^{2}sT_{p}(\vec{b}-\vec{s})e^{-\sigma_{\rm eff}A_{t}T_{t}(\vec{s})},\label{prob15} 
\end{equation}

where $\vec{b}$ is the impact parameter of the nucleus-nucleus
collision, $T_{p}(b)$ and $T_{t}(b)$ are the thickness functions
($T(b)=\int dz\rho (b,z),$ $\int d^{3}x\rho (x)=1$) of projectile and
target, respectively. The straight line geometry is certainly
justified for the through-going nucleons. The use of ``frozen''
nucleons and the neglect of any other degrees of freedom, like mesons,
need justification. We consider a target nucleon and estimate the
time $\Delta t$ in its rest system which has elapsed between the
arrival times of the first and the last projectile nucleons: $\Delta
t\leq 2R_{p}/\gamma_{p}$ where $2R_{p}\simeq 7$ fm is the diameter of
the Si projectile and $\gamma_{p}\simeq 16$ is the Lorenz factor
for the experiment under consideration. The time $\Delta t\approx 0.5$ 
fm/c is rather short: (i) The target nucleon has not moved
significantly in space (frozen approximation is good). (ii)
According to the uncertainty principle the intrinsic excitation energy
is uncertain by $\Delta E\geq (\Delta t)^{-1}\approx 0.4$ GeV and does
not allow the assignment of a definite excited state. (iii) Secondary
hadrons are not yet formed since typical formation times are of order
1 fm/c.

\begin{figure}[t]
  \begin{center}
      \epsfig{file=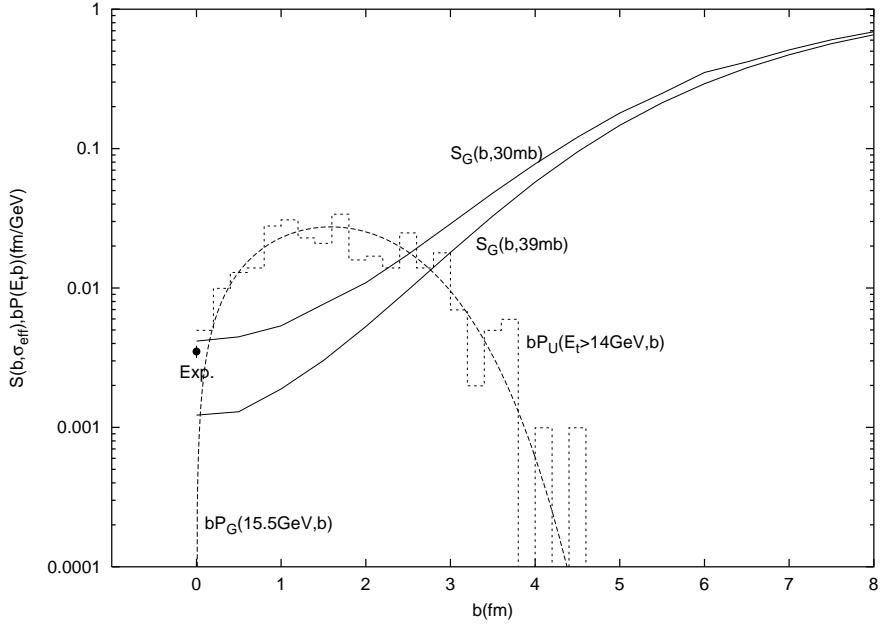,width=12cm}
  \end{center}
\caption{Survival probabilities $S_{G}(b,\sigma_{\rm eff})$ of a
beam rapidity nucleon calculated in Glauber approximation for
$\sigma_{\rm eff}=\sigma_{in}^{NN}=30$ mb and $\sigma_{\rm
eff}=\sigma_{tot}^{NN}=39$ mb (solid lines), and the experimental
value $S_{exp}$ for the survival probability for $E_{t}^{c}=15.5$ 
GeV. The correlation function $P(E_{t},b)$ between transverse energy
$E_{t}$ and impact parameter b is (i) calculated in Glauber
approximation, $P_{G}(E_{t},b)$ (continuous line), and (ii) deduced
from the results of UrQMD, $P_{U}(E_{t},b)$ (histogram), both plotted as 
$bP$ in order to represent d$\sigma/$d$b$.}
\end{figure}

The value of the transparency $S$ in Eq.~(\ref{prob15}) depends on the
effective cross section $\sigma_{\rm eff}$. Using Saxon-Woods
parameterizations for the densities $\rho_{t}$ and $\rho_{p}$ with
the surface thickness $a=0.52$ fm for all nuclei and the half-density
radius $r_{A}$ such that the root-mean-square radius of the nucleus
equals the charge radius \cite{3}, the survival probability
$S(b,\sigma_{\rm eff})$ is calculated for two values
$\sigma_{in}^{NN}=30$ and $\sigma^{NN}_{tot}=39$ mb which correspond
to the inelastic NN cross section and the total one at this energy,
respectively. The results are displayed in Fig.~1. The
experimental point, measured at transverse energy $E_{t}^{c}=15.5$ GeV
is also shown in the figure. It is obtained from the measured mean
multiplicity $\langle M_{c}\rangle$ of beam rapidity protons by

\begin{equation}
S_{exp}(E_{t})=\langle M_{c}\rangle (E_{t})/Z,
\end{equation}

where $Z=14$ is the number of protons in Si. We have assumed -- as
the authors of the experiment do -- that the highest $E^{c}_{t} $-bin
corresponds to a central collision which is assigned a value
$b=0$. Then the experimental value is close to the curve $\sigma_{\rm
eff}=30$ mb. In fact the equation $S_{G}(b=0,\sigma_{\rm
eff})=S_{exp}(E_{t}^{c}=15.5$ GeV) leads to a value $\sigma_{\rm
eff}=31.1\pm 0.7$ mb. This result reproduces a similar calculation
using uniform density distributions by the E814 collaboration
\cite{1}, who have concluded that $\sigma_{\rm eff}=\sigma_{in}^{NN}$
within error bars and no anomaly being visible.

The crucial step in the argument is the assignment of $b=0$ to the
highest $E_{t}$-bin.  Actually, a given value of $E_{t}$ selects a
distribution of impact parameters only within a band $\Delta b(E_{t})$
which is quite large.  We study the relation between $E_{t}$ and $b$
and $\Delta b$ in the form of a probability distribution$ P(E_{t},b)$,
which gives the probability that values of $\vec b$ contribute to events
with a given value of $E_{t}$. We normalize it as $\int
dE_{t}P(E_{t},b)=1$. With this function and the differential inelastic
heavy ion cross section d$^2\sigma_{in}/d^{2}b $, one can obtain the
dependence of the survival function $S(E_{t},\sigma_{\rm eff})$ as a 
function of the transverse energy

\begin{equation}
\label{pet}
S(E_{t},\sigma_{\rm eff})=\frac{\int d^{2}bS(b,\sigma_{\rm eff})P(E_{t},b)d\sigma_{in}/d^{2}b}{d\sigma_{in}/dE_{t}},
\end{equation}
 where

\begin{equation}
\label{s_et}
\frac{d\sigma_{in}}{dE_{t}}=\int d^{2}bP(E_{t},b)\frac{d\sigma_{in}}{d^{2}b}.
\end{equation}

We will use Eq.~(\ref{s_et}) to determine $ P(E_{t},b)$ from a
comparison with the measured transverse energy distribution $d\sigma
_{in}/dE_{t}$.  With a reasonable Gaussian parameterization of $
P_G(E_{t},b)$ for the Glauber calculation (or with $P_U(E_{t},b)$
from a UrQMD calculation, see below), $S(E_{t},\sigma_{\rm eff})$ can
be determined from Eqs.~1 and 3 without ambiguity.  We discuss our
parameterizations.

The inelastic cross section d$^2\sigma_{in}/d^{2}b$ is taken from the
folding model

\begin{equation}
\label{s_b}
\frac{d\sigma_{in}}{d^{2}b}(b)=1-exp[-\sigma ^{NN}_{in}A_{p}A_{t}\int d^{2}sT_{p}(\vec{b}-\vec{s})T_{t}(s)].
\end{equation}

 The parameterization of $P(E_{t},b)$ is more model dependent, and we
calculate it in two ways: (i) in the formalism of Glauber theory, and
call it $P_{G}(E_{t},b)$, and (ii) deduce it from results of the
cascade calculation, $P_{U}(E_{t},b)$. To calculate $P_{G}$ we assume
a Gaussian parameterization

\begin{equation}
\label{p_et_b}
P_{G}(E_{t},b)=\frac{1}{\sqrt{2\pi \sigma ^{2}_{t}(b)}}exp\{-\frac{[E_{t}-E_{t}(b)]^{2}}{2\sigma ^{2}_{t}(b)}\},
\end{equation}

which satisfies the normalization condition. We make the usual
assumptions \cite{4,5} for the functions $E_{t}(b)$ and $\sigma_{t}(b)$

\begin{equation}
\label{E_b}
E_{t}(b)=N(b)\epsilon_{0},
\end{equation}

\begin{equation}
\label{st_b}
\sigma ^{2}_{t}(b)=N(b)\epsilon_{0}^{2}\omega.
\end{equation}

Here $N(b)$ is calculated in the ``collision model'' \cite{4} 

\begin{equation}
\label{N_b}
N(b)=\sigma_{in}^{NN}A_{t}A_{p}\int d^{2}bT_{p}(\vec{s})T_{t}(\vec{b}-\vec{s}),
\end{equation}

which equals the mean number of NN collisions in a projectile-target
interaction with impact parameter $b$. The proportionality constant
$\epsilon_{0}$ between the observed transverse energy $E_{t}$ and
$N(b)$ depends on the dynamics of hadron production, but also on the
experimental set up (chosen rapidity interval and acceptance of
counter). It will be a fit parameter. The ansatz
$\sigma^{2}_{t}\propto N(b)$ corresponds to the hypothesis that the
fluctuations in $N(b)$ are of statistical origin. It is known,
however, that the proportionality constant $\omega$ in
Eq.~(\ref{st_b}) depends strongly on the rapidity interval of the
accepted particles \cite{4}. The reason is not clear \cite{6}. We take
$\omega$ as a free parameter.

\begin{figure}
  \begin{center}
     \epsfig{file=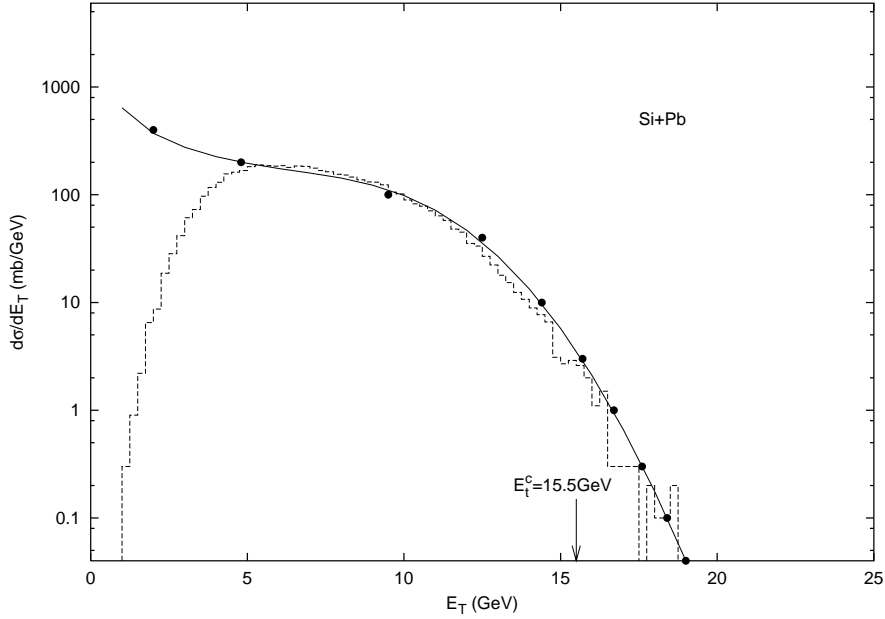,width=12cm}
  \end{center}
\caption{Transverse energy distributions in the rapidity range
$-0.5<\eta<0.8$ for Si+Pb collisions at 14.6 GeV/nucleon. Data
points (dots) are taken from \cite{7}. The solid line is calculated
within the collision model with parameters
$\epsilon_{0}$, $\omega$ determined by a fit to the Pb data. The
UrQMD calculation (histogram) contains a simulation of the leakage of
the TCAL counter as explained in the text.}
\end{figure}

Using expressions (\ref{s_b}) to (\ref{st_b}) we have calculated
$d\sigma_{in}/dE_{t}$ and have varied the parameters $\epsilon_{0}$
and $\omega$ until the data for Si-Pb are fitted. The result is
shown in Fig.~2 with $\sigma_{in}^{NN}=30$ mb, $\epsilon_{0}=0.067$
GeV, $\omega =7.8$.  A calculation of $N(b)$ in
Eqs.~(\ref{E_b}-\ref{st_b}) within the wounded-nucleon model \cite{5}
gives a similarly good fit to the data for $d\sigma_{in}/dE_{t}$.

According to Eq.~(\ref{pet}), $P(E_{t},b)$ determines the integration
region in impact parameter $b$, which contributes to the integral for
a given value of $E_{t}$. Fig.~1 shows the distribution
$bP_{G}(E_{t},b)$ for a central collision ($E_{t}^{c}=15.5$ GeV)
derived in the Glauber formalism, and the corresponding one,
$bP_{U}(E_{t}^{c},b)$, calculated (see below) with the help of the
code UrQMD \cite{8}. Both distributions agree within statistics, with
a peak around 1.7 fm with a half width of about 2.5 fm. Using the
functions $S_{G}(b,\sigma_{\rm eff})$ and $P_{G}(E_{t}^{c},b)$ as
shown in Fig.~1, the calculated value of $S_{G}(E_{t}^{c}$,
$\sigma_{\rm eff})$ gets contributions from a considerable range of
values of $b$. We find

\begin{equation}
S(E_{t}^{c},\sigma_{\rm eff}=30 {\rm mb})/S_{exp}=3.47
\end{equation}

The calculated transparency is too large by a factor of three.. Should
one rather use $\sigma_{\rm eff}=\sigma_{tot}^{NN}$ instead? We think
yes: Although elastic scattering changes the rapidity of the
projectile nucleon only marginally, it increases the transverse
momentum of a nucleon, and an elastically scattered nucleon is not
accepted within the experimental window $p_{t}<0.3$ GeV with a large
probability. The results of the Glauber-type calculations are compared
with experiment in Fig.~4. We discuss this figure after we have
described the second approach.

The UrQMD calculations presented in this paper are based on the
standard version 1.1 \cite{8}. As mentioned in \cite{1}, due to leakage
in the TCAL \cite{7}, there is less than half of the $E_{t}$
visible. This has two consequences:

\begin{figure}
  \begin{center}
     \epsfig{file=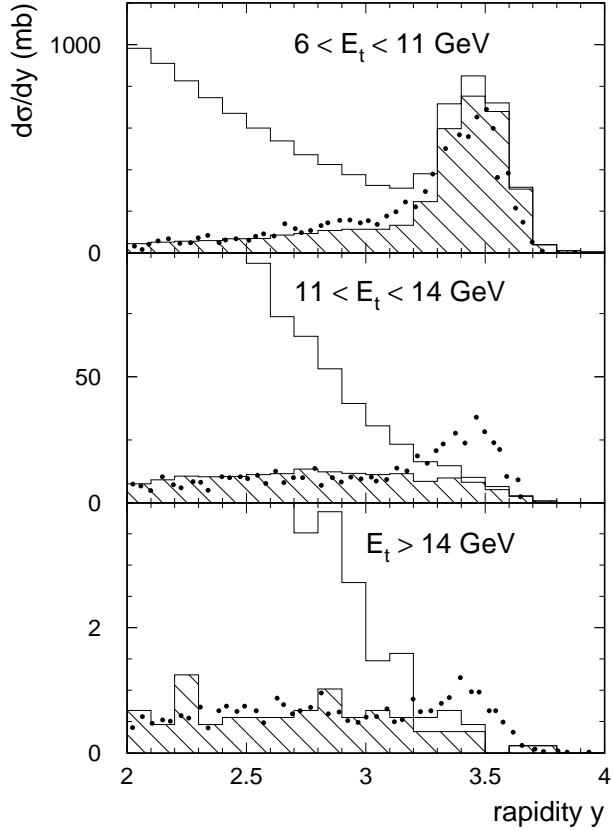,width=8cm,
bbllx=60,bburx=550,bblly=82,bbury=755}
  \end{center}
\caption{The rapidity distributions $d\sigma /dy$ for protons and
neutrons with $p_{t}<0.3$ GeV/c for three intervals of transverse
energy $E_{t}$. The data are indicated by dots, the calculations using
UrQMD are represented by histograms. In each figure the histogram with
higher values corresponds to ideal acceptance, the lower, hatched
histogram is obtained with E814 acceptance cuts.  Beam rapidity is at
$y=3.44$, where the data show a clear peak while the calculation does
not show one for the two higher $E_{t}$ bins.}
\end{figure}

(i) The $E_t$ scale of the data cannot be directly compared with that
of the cascade calculation.  (ii) The data have additional
fluctuations, visible as a flatter slope of the E814
$E_t$-distribution beyond the knee. Simulating the leakage in the
Monte Carlo by simply selecting only 37\% of the particles hitting the
TCAL fixes both problems.  The resulting $E_t$-distribution agrees
quite well with the data (Fig.~2). In order to compare to the
$E_t$-selected proton rapidity distributions of E814, we have used the
absolute cross section given there -- 593, 102 and 6.9 mb -- for the
three centrality bins shown in Fig.~3. The mean impact parameters for
these bins are 3.6, 2.1, and 1.6 fm, resp., in UrQMD.

Fig.~3 shows the rapidity distributions for protons + neutrons for the
three $E_{t}$ bins. The data points are shown by the dots, while the
two histograms represent the results of the UrQMD calculation with an
upper cut in $p_{t}$ at 0.3 GeV/c. The two histograms show results
assuming either perfect acceptance or taking the E814 experimental
acceptance cut into account (hatched area). The two histograms are
rather similar in the region of beam rapidity ($y_B=3.44$). While
calculations agree with the data in the interval $6<E_{t}<11$ GeV,
there is a clear discrepancy in the region around beam rapidity for
the two other $E_{t}$ intervals (middle and lower picture): The
calculations do not show any peak, while the data do. 

\begin{figure}
  \begin{center}
     \epsfig{file=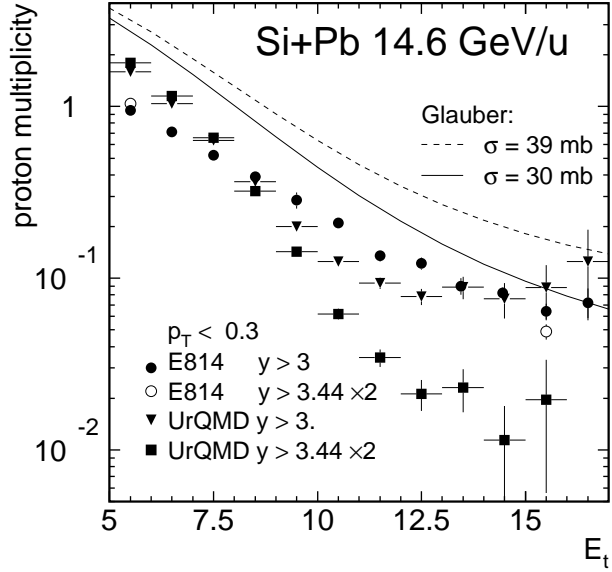,width=8cm,
bbllx=112,bburx=447,bblly=264,bbury=582}
  \end{center}
\caption{Comparison of experimental and theoretical results for the
multiplicity of beam rapidity protons as a function of transverse
energy $E_{t}$ for the reaction Si+Pb at 14.6 GeV/nucleon. Two
methods are employed to characterize ``beam rapidity protons'' in
experiment and in the output of UrQMD: (i) protons with rapidity $y>3$
(black circles for the data and triangles for UrQMD) and (ii)
multiplicity of protons with $y>y_B (3.44)$ multiplied by a factor 2 (open
circles for the data and squares for UrQMD). The dashed and solid
lines represent Glauber calculations with $\sigma_{\rm
eff}=\sigma_{in}=30$ mb and $\sigma_{\rm eff}=\sigma_{tot}=39$ mb,
respectively.}
\end{figure}

Fig.~4 shows a comparison between experiment and our two calculations
for the multiplicity of beam rapidity protons. In the analysis of the
data and in the output of UrQMD, two different criteria are employed
to characterize ``beam rapidity protons'': (i) all protons with $y>3$
(black dots for the experiment and triangles for the UrQMD
calculation); (ii) twice the number of protons with $y_B>3.44$ -
assuming the distribution to be symmetric around $y_B$ (open circles
for the experiment and black squares for the UrQMD calculation). While
the two criteria do not give very different results for the
experiment, they lead to very different values for UrQMD up to a
factor of five smaller for the second method. This reflects the fact
already observed in Fig.~3 that UrQMD does not produce a peak at beam
rapidity for the more central values of $E_{t}$.  It is therefore not
meaningful to compare data and results from UrQMD for $E_{t}>10$
GeV. For values of $E_{t}<10$ GeV calculation and data agree to better
than a factor of two -- the calculations are above the data for
$E_{t}=5$ GeV and below the data at $E_{t}=10$ GeV. The result of the
Glauber calculations (solid curve for $\sigma_{\rm
eff}=\sigma_{tot}=39$ mb, dashed curve $\sigma_{\rm
eff}=\sigma_{in}=30$ mb) are nearly always above the data. The solid
curve agrees within 30-50\% for central values of $E_{t}$, but
agreement worsens towards smaller values of $E_{t}$. At $E_{t}=5$ GeV
the Glauber calculation is a factor 3 higher than the data. Note that
in this region also the results of UrQMD lie above the data by a
factor of 1.5. These discrepancies may partly have the following
origin: Small values of $E_{t}$ correspond to peripheral reactions
with several nucleons being spectator particles. All spectator
nucleons are included in the calculations; spectator nucleons bound in
deuterons, alpha particles etc., however, are not contained in the
data. 

In summary, the two theoretical approaches, Glauber theory and UrQMD,
applied to the transverse energy dependence of the nuclear
transparency, have yielded only a semi-quantitative understanding of
the data. At small values of $E_t$ (peripheral reactions) calculations
are systematically above the data. In the interesting regime of
central collisions, Glauber theory predicts values of 30 to 50\% above
the data, while UrQMD shows clearly no peak of beam-rapidity particles
so that the results are only an upper limit.

Although the theoretical situation is still far from being satisfactory
the analysis of beam-rapidity particles in heavy ion collisions may
potentially be a good way to investigate the question of effective cross
sections of woun\-ded nucleons. 

\begin{ack}
 
We thank P. Braun-Munzinger, S. Kahana, B. Kope\-lio\-vich and
J. Stachel for several discussions. H.M.\ Ding is grateful to the
Institut f\"ur Theoretische Physik, Universit\"at Heidelberg for its
hospitality and is grateful for a fellowship of the government of the
People's Republic of China. This research is partly supported by the
BMFT under grant O6HD856.

\end{ack}

\end{document}